\documentclass[a4paper,12pt]{article}

\usepackage{amsmath,amssymb}
\usepackage{graphicx}

\usepackage{pnas}  
\usepackage[round,numbers,sort&compress]{natbib}

\usepackage[small,bf]{caption2}
\usepackage{booktabs}

\newcommand{\url}[1]{{#1}}

\newsavebox{\junk}
\newcommand{\citesilent}[1]{{\sbox{\junk}{#1}}}

\begin{document}

\title{Liquid-vapor oscillations of water in hydrophobic nanopores}
\author{Oliver Beckstein \and Mark S.\ P.\ Sansom$^{*}$}
\date{\normalsize
  \vspace{1cm}
  \noindent
  \raggedright
  Department of Biochemistry,\\
  University of Oxford,\\
  South Parks Road,\\
  Oxford.\\
  OX1 3QU,\\
  U.K.\\[1cm]
  \noindent
  $^{*}$To whom correspondence should be addressed at\\
  \begin{tabular}{ll}
    email: & mark@biop.ox.ac.uk\\
    Tel:   & +44--1865--275371\\
    Fax:   & +44--1865--275182
  \end{tabular}
  \vfill
}
\maketitle
\newpage

\begin{abstract}
  \sffamily\noindent Water plays a key role in biological membrane transport.
  In ion channels and water-conducting pores (aquaporins), one dimensional
  confinement in conjunction with strong surface effects changes the physical
  behavior of water.  In molecular dynamics simulations of water in short
  (0.8~nm) hydrophobic pores the water density in the pore fluctuates on a
  nanosecond time scale.  In long simulations (460~ns in total) at pore radii
  ranging from 0.35~nm to 1.0~nm we quantify the kinetics of oscillations
  between a liquid-filled and a vapor-filled pore. This behavior can be
  explained as capillary evaporation alternating with capillary condensation,
  driven by pressure fluctuations in the water outside the pore. The free
  energy difference between the two states depends linearly on the radius. The
  free energy landscape shows how a metastable liquid state gradually develops
  with increasing radius.  For radii larger than ca.~0.55~nm it becomes the
  globally stable state and the vapor state vanishes. One dimensional
  confinement affects the dynamic behavior of the water molecules and
  increases the self diffusion by a factor of two to three compared to bulk
  water. Permeabilities for the narrow pores are of the same order of
  magnitude as for biological water pores. Water flow is not continuous but
  occurs in bursts. Our results suggest that simulations aimed at collective
  phenomena such as hydrophobic effects may require simulation times longer
  than 50~ns.  For water in confined geometries, it is not possible to
  extrapolate from bulk or short time behavior to longer time scales.
\end{abstract}

\section{Introduction}
\label{sec:introduction}

Channel and transporter proteins control flow of water, ions and other solutes
across cell membranes.  In recent years several channel and pore structures
have been solved at near atomic resolution
\citep{Doy98,Cha98,Fu00,Sui01,Bass02,Dutzler02} which together with three
decades of physiological data \citep{Hille01} and theoretical and simulation
approaches \citep{Tieleman01} allow us to describe transport of ions, water or
other small molecules at a molecular level.  Water plays a special role here:
it either solvates the inner surfaces of the pore and the permeators (for
example, ions and small molecules like glycerol), or it is the permeant
species itself as in the aquaporin family of water pores
\citep{Yang97,Pohl01,Fujiyoshi02R} or in the bacterial peptide channel
gramicidin A (gA), whose water transport properties are well studied
\citep{Pohl00,Saparov00,DeGroot02}. Thus, a better characterization of the
behavior of water would improve our understanding of the biological function
of a wide range of transporters. The remarkable water transport properties of
aquaporins---water is conducted through a long (ca.\ 2~nm) and narrow (ca.\ 
0.3~nm diameter) pore at bulk diffusion rates while at the same time protons
are strongly selected against---are the topic of recent simulation studies
\citep{Tajkhorshid02,DeGroot01}.

The shape and dimensions of biological pores and the nature of the pore lining
atoms are recognized as major determinants of function. How the behavior of
water depends on these factors is far from understood \citep{Finkelstein87}.
Water is not a simple liquid due to its strong hydrogen bond network. When
confined to narrow geometries like slits or pores it displays an even more
diverse behavior than already shown in its bulk state
\citep{Christenson01,Gelb99}.

A biological channel can be crudely approximated as a ``hole'' through a
membrane. Earlier molecular dynamics (MD) simulations showed pronounced
layering effects and a marked decrease in water self diffusion in infinite
hydrophobic pore models \citep{Lyn96,All99}.  Recently, water in finite narrow
hydrophobic pores was observed to exhibit a distinct two-state behavior. The
cavity is either filled with water at approximately bulk density (liquid-like)
or it is almost completely empty (vapor-like) \citep{Hum01,Beckstein01}.
Similar behavior was seen in Gibbs ensemble Monte Carlo simulations (GEMC) in
spherical \citep{Bro00} and cylindrical pores \citep{Brovchenko02}.

In our previous simulations \citep{Beckstein01} we explored model pores of the
dimensions of the gating region of the nicotinic acetylcholine receptor nAChR
\citep{Unw00}. Hydrophobic pores of radius $R \geq 0.7$~nm were filled during
the whole simulation time (up to 6~ns) whereas narrow ones ($R \leq 0.4$~nm)
were permanently empty.  Changing the pore lining from a hydrophobic surface
to a more hydrophilic (polar) one rendered even narrow pores water---and
presumably ion---conducting. At intermediate radii ($0.4$~nm $< R < 0.7$~nm)
the pore-water system was very sensitive to small changes in radius or
character of the pore lining.  In a biological context, a structure close to
the transition radius would confer the highest susceptibility to small
conformational rearrangements (i.e.\ gating) of a channel.

\begin{figure}
    \hspace*{-10mm}
    \includegraphics[clip=,height=42mm]{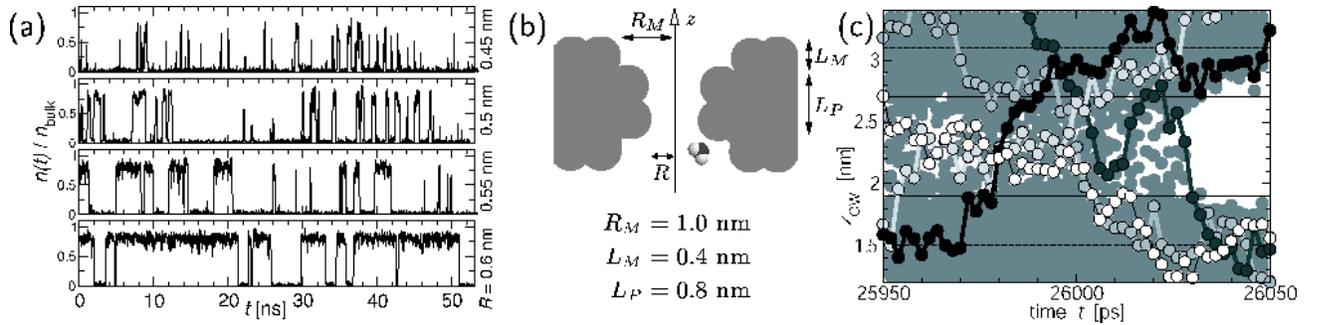}
    \caption{\label{fig:states}%
      (\emph{a}) Oscillating water density in model pores of increasing pore
      radius $R$.  The water density $n(t)$ (in units of the bulk water
      density $n_{\text{bulk}}$) over the simulation time shows strong
      fluctuations on a greater than ns time scale ($50$~ps moving average
      smoothing).  Two distinctive states are visible: \emph{open} at
      approximately $n_{\text{bulk}}$ (liquid water), and \emph{closed} with
      very few or no water in the pore (water vapor).
      \label{fig:geometry}
      (\emph{b}) The pore model consists of methane
      pseudo atoms of van der Waals radius $0.195$~nm.  A water molecule is
      drawn to scale.
      \label{fig:wcrossing}%
      (\emph{c}) Permeant water molecules in a $R=0.55$~nm pore as it
      switches from the 
      open to the closed state. $z$-coordinates of the water
      oxygen atoms are drawn every 2~ps. The mouth and pore region are
      indicated 
      by horizontal broken and solid lines. Five trajectories are shown
      explicitly. The white water molecule permeates the pore within 54~ps
      whereas the black one only requires about 10~ps.}
\end{figure}

We have extended these simulations to beyond 50~ns in order to explore the
longer timescale behavior of the water-pore system. Starting from the observed
oscillations in water density (Fig.~\ref{fig:states}\emph{a}) we analyze the
kinetics and the free energy of the system. We compare the water transport
properties of the pores to experimental and theoretical data.

\section{Methods}
\label{sec:methods}

\subsection{Model}
\label{sec:model}

The pore model was designed to mimic the dimensions of a biological pore
[e.g., the gate region of nAChR \citep{Unw00}], whilst retaining the
tractability of a simple model.  Cylindrical pores of finite length were
constructed from concentric rings of pseudo atoms
(Fig.~\ref{fig:geometry}\emph{b}).  These hydrophobic pseudo atoms have the
characteristics of methane molecules, i.e.\ they are uncharged Lennard-Jones
(LJ) spheres with a van der Waals radius of 0.195~nm.  The pore consists of
two mouth regions (radius $R_M = 1$~nm, length $L_M = 0.4$~nm) and an inner
pore region ($L_P = 0.8$~nm) of radius $0.35\ \text{nm} \leq R \leq 1.0\ 
\text{nm}$, which is the minimum van der Waals radius of the cavity. The
centers of the pore lining atoms are placed on circles of radius $R+0.195$~nm.
The model was embedded in a membrane mimetic, a slab of pseudo atoms of
thickness ca.\ 1.5~nm or 1.9~nm.  Pseudo atoms were harmonically restrained to
their initial position with a force constant of
$1000$~kJ\,mol$^{-1}$\,nm$^{-2}$, resulting in positional fluctuations of
ca.~0.1~nm, comparable to those of pore lining atoms in membrane protein
simulations although this does not allow for global collective motions as in
real proteins.

\subsection{Simulation Details}
\label{sec:sim}

MD simulations were performed with \textsc{gromacs} v3.0.5 \citep{Lindahl01}
and the SPC water model \citep{Hermans84}. The LJ-parameters for the
interaction between a methane molecule and the water oxygen are
$\epsilon_{\text{CO}}=0.906493$~kJ~mol$^{-1}$ and
$\sigma_{\text{CO}}=0.342692$~nm from the \textsc{gromacs} force field. The
integration time step was 2~fs and coordinates were saved every 2~ps. With
periodic boundary conditions, long range electrostatic interactions were
computed with a particle mesh Ewald method [real space cutoff 1~nm, grid
spacing 0.15~nm, 4th order interpolation \citep{Dar93}] while the short ranged
van der Waals forces were calculated within a radius of 1~nm. The neighbor
list (radius 1~nm) was updated every 10 steps.

Weak coupling algorithms \citep{Ber84} were used to simulate at constant
temperature ($T=300$~K, time constant 0.1~ps) and pressure ($P=1$~bar,
compressibility $4.5\times10^{-5}$~bar$^{-1}$, time constant 1~ps) with the
$x$ and $y$ dimensions of the simulation cell held fixed at 6~nm (or 3.9~nm
for the 80~ns simulation of the $R=0.35$~nm pore). The length in $z$ was
4.6~nm in both cases (ensuring bulk-like water behavior far from the membrane
mimetic).

The large (small) system contained about 700 (300) methane pseudo atoms and
4000 (1500) SPC water molecules.  Simulation times $T_{\text{sim}}$ ranged
from 52~ns to 80~ns (altogether 460~ns). Bulk properties of SPC water were
obtained from simulations in a cubic cell of length 3~nm (895 molecules) with
isotropic pressure coupling at 300~K and 1~bar for 5~ns.

\subsection{Analysis}
\label{sec:analysis}

\paragraph{Time courses and density}

For the density time courses (Fig.~\ref{fig:states}\emph{a}) the pore
occupancy $N(t)$, i.e.\ the number of water molecules within the pore cavity
(a cylinder of height $L_{P}=0.8$~nm containing the pore lining atoms) was
counted.  The density $n(t)$ is given by $N(t)$ divided by the
water-accessible pore volume $V=L_{P}\,\pi R_{\text{eff}}^{2}$ and normalized
to the bulk density of SPC water at 300~K and 1~bar
($n_{\text{bulk}}=53.67\pm0.03$~mol\,l$^{-1}$).  The effective pore radius for
all pores is $R_{\text{eff}}=R-\delta R$.  Choosing $\delta R=0.03$~nm fixes
the density $\langle N \rangle/V$ in the (most bulk-like) $R=1.0$~nm-pore at
the value calculated from the radial density,
$R_{0}^{-1}\!\!\int_{0}^{R_{0}}\!\!\!\,n(r)\,dr$, where $R_{0}=1.05$~nm is the
radius at which $n(r)$ vanishes.

The density $n(\mathbf{r})$ was determined on a grid of cubic cells with
spacing 0.05~nm.  Two- and one-dimensional densities were computed by
integrating out the appropriate coordinate(s). A probabilistic interpretation
of $n(\mathbf{r})$ leads to the definition of the potential of mean force
(PMF) of a water molecule $\beta F(\mathbf{r})=-\ln
[n(\mathbf{r})/n_{\text{bulk}}]$ with $\beta^{-1}=k_{B}T$, via
Boltzmann-sampling of states.

\paragraph{Free energy density and chemical potential}

The Helmholtz free energy as a function of the pore occupancy $N$ at constant
$T=300$~K for a given pore with volume $V$ was calculated from the probability
distribution $p(N)$ of the occupancy as $\beta F(T,V,N) = -\ln p(N)$ and
transformed into a free energy density $f(T,n)=F/V$.  A fourth order
polynomial in $n$ was least-square fitted to $\beta f(T,n)$. The chemical
potential $\mu(T,n) = \partial f(T,n)/\partial n$ was calculated as the
analytical derivative of the polynomial.  

The $\beta f$ curves obtained for different radii $R$ from the simulations are
only determined within an unknown additive constant $f_{0}(T;R)$ but a
thermodynamic argument shows that all these curves  coincide at $n=0$: For
$n=0$ no water is in the pore, so the free energy differential is simply
$dF=-S\,dT + 2\gamma\, dA$ with the constant surface tension of the vacuum
(inside the pore)-water (outside) interface of area $A=\pi R^{2}$.  At
constant $T$ this implies $F(R)=2\gamma\,A + \text{const}(T)$, so that the
free energy density $f(T,n=0;R)=F(R)/V=2\gamma\,A/(L\,A)=2\gamma/L$ of a pore
with radius $R$ and length $L$ is independent of $R$. Hence, all free energy
density curves necessarily coincide at $n=0$ and $f_{0}$ is a function of $T$
only.

\paragraph{Kinetics}

The time series $n(t)/n_{\text{bulk}}$ of the water density in the pore was
analyzed in the spirit of single channel recordings \citep{Sakmann83} by
detecting open (high-density; in the following denoted by a subscript $o$) and
closed (approximately zero density; subscript $c$) pore states, using a
Schmitt-trigger with an upper threshold of 0.65 and a lower threshold of
0.15\,.  A characteristic measure for the behavior of these pores is the
\emph{openness} $\langle\omega\rangle=T_{o}/T_{\text{sim}}$, i.e.\ the
probability for the pore being in the open state \citep{Beckstein01} with
errors estimated from a block-averaging procedure \citep{Lindahl01}.  The
distribution of the lifetimes $t_{\alpha}$ of state $\alpha = \{o, c\}$ are
exponentials $\tau_{\alpha}^{-1} e^{-t_{\alpha}/\tau_{\alpha}}$ (data not
shown). The maximum-likelihood estimator for the characteristic times
$\tau_{o}$ and $\tau_{c}$ is the mean $\tau_{\alpha} = \langle t_{\alpha}
\rangle$.

The free energy difference between closed and open state, $\Delta F = F_{c} -
F_{o}$, can be calculated in two ways. Firstly, we obtained it from the
equilibrium constant $K=T_{c}/T_{o} = (T_{\text{sim}}-T_{o})/T_{o} =
\langle\omega\rangle^{-1}-1$ of the two-state system as $\beta\, \Delta
F_{\text{kin}} = -\ln K$.  Secondly, $\Delta F$ was determined from $p(N)$ as
the ratio between the probability that the pore is in the
closed state and the probability for the open state:
$\beta\, \Delta F_{\text{eq}} = -\ln P_{c}/P_{o}
       = -\ln \sum_{N \leq N^{\ddagger}}p(N)/
                   \sum_{N > N^{\ddagger}}p(N)$.
The definition of state used here is independent of the kinetic analysis. It
only depends on $N^{\ddagger}$, the pore occupancy in the transition state,
the state of lowest probability between the two probability maxima that define
the closed and open state. The relationship involving $K$ can be inverted to
describe the openness in terms of $\Delta F(R)$,
$\langle\omega(R)\rangle=\bigl(1+\exp[-\beta\,\Delta F(R)]\bigr)^{-1}$.

\paragraph{Dynamics}

The three components of the self-diffusion coefficient were calculated from
the Einstein relations $\bigl\langle \bigl(x_{i}(t) - x_{i}(t_{0})\bigr)^{2}
\bigr\rangle = 2 D_{i}\, (t-t_{0})$. The simulation box was stratified
perpendicular to the pore axis with the central layer containing the pore.
During $T_{\text{sim}}$ the mean square deviation (msd) of water molecules in
a given layer was accumulated for 10~ps and after discarding the first 2~ps, a
straight line was fit to the msd to obtain $D_{i}$. These diffusion
coefficients were averaged in each layer for the final result.

The current density (flux per area) was calculated as $j_{0}=\Phi_{0}/A$
from the equilibrium flux $\Phi_{0}=M/T_{\text{sim}}$ with the total number
of permeant water molecules $M$ and the effective pore cross section $A=\pi
R_{\text{eff}}^2$ for pores or $A=L_x L_y$ for the bulk case, i.e.\ a
simulation box of water with periodic boundary conditions.  Permeant water
molecules were defined as those whose pore entrance and exit $z$-coordinate
differed. In addition, distributions of permeation times were computed.

\section{Results and Discussion}
\label{sec:results}

The water density in the pore cavity oscillates between an almost empty
(closed) and filled (open) state (Fig.~\ref{fig:states}\emph{a}). We refer to
the water-filled pore state as open because such a pore environment would
favorably solvate an ion and conceivably allow its permeation.  Conversely, we
assume that a pore that cannot sustain water at liquid densities will present
a significant energetic barrier to an ion. As shown in
Fig.~\ref{fig:wcrossing}\emph{c}, water molecules can pass each other and
often permeate the pore in opposite directions simultaneously.

\begin{figure}
  \centering
    \includegraphics[clip]{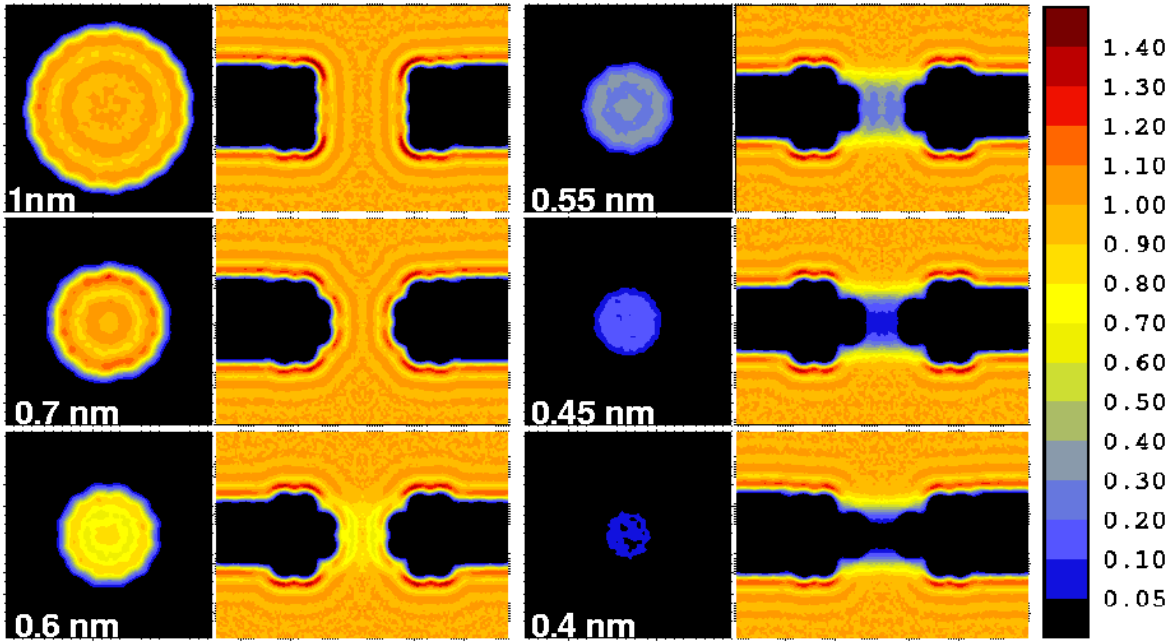}
    \caption{\label{fig:densities}%
      Water density in hydrophobic pores with radii ranging from 1.0~nm to
      0.4~nm.  Left column in each panel: density $z$-averaged over the length
      of the pore.  Right column: radially averaged density.  The density is
      in units of SPC bulk water at 300~K and 1~bar [plots prepared with
      xfarbe~2.5 \citep{Preusser89}].}
\end{figure}

Even though the oscillating behavior was already suggested by earlier $1$~ns
simulations \citep{Beckstein01} only at these longer times do clear patterns
emerge.  The characteristics of the pore-water system change substantially
with the pore radius. The oscillations (Fig.~\ref{fig:states}\emph{a}) depend
strongly on the radius. The water density (Fig.~\ref{fig:densities})
\citesilent{\citep{Preusser89}} shows large pores to be water-filled and
strongly layered at bulk density. With decreasing radius the average density
is reduced due to longer closed states even though layer structures remain.
The narrowest pores appear almost void of water.

\begin{figure}
  \centering
  \hspace*{-7mm}
  \includegraphics[clip]{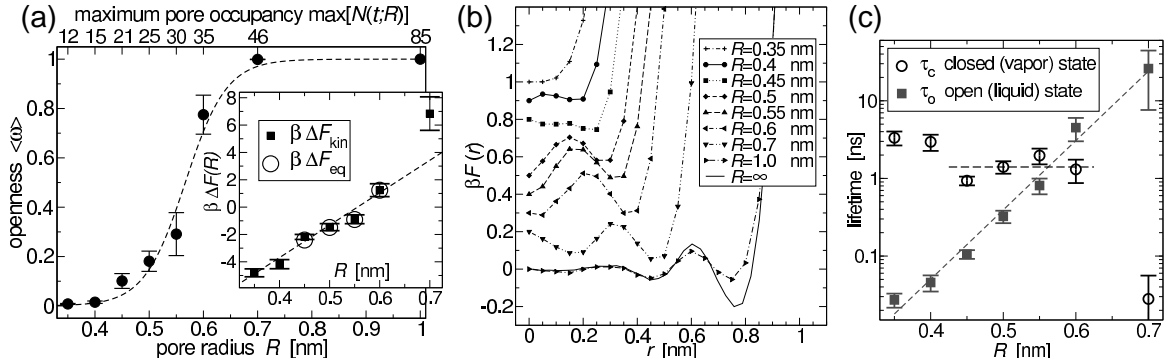}
  \caption{\label{fig:openness}%
    (\emph{a}) Openness $\langle\omega(R)\rangle$ of hydrophobic pores and
    free energy difference $\Delta F(R)$ between states (inset). Wide pores
    are permanently water-filled ($\langle\omega\rangle=1$) whereas narrow
    ones are predominantly empty ($\langle\omega\rangle\approx 0$). The broken
    line is the function $\bigl(1+\exp[-\beta\,\Delta
    F_{\text{eq}}(R)]\bigr)^{-1}$, with $\Delta F_{\text{eq}}(R)$
    determined independently of $\langle\omega(R)\rangle$.  $\Delta F(R)$
    appears to be a linear function of $R$, regardless if estimated from the
    kinetics ($\Delta F_{\text{kin}}$) or the equilibrium probability
    distribution of the pore occupancy ($\Delta F_{\text{eq}}$).
    \label{fig:radpmf}
    (\emph{b}) Radial potential of mean force of water $F(r)$. Very narrow
    pores show a relatively featureless PMF, consistent with a predominantly
    vapor-like state.  For larger pore radii, the liquid state dominates. The
    PMF of the 1~nm pore is very similar to the one of water near a planar
    hydrophobic slab ($R=\infty$). PMFs are drawn with arbitrary offsets.
    \label{fig:kinetics}
    (\emph{c}) Kinetics \emph{open}$\rightleftharpoons$\emph{closed}. The
    average lifetime of the open state $\tau_{o}$ depends on the radius
    exponentially whereas $\tau_{c}$ is approximately constant in the
    two-state region (cf.\ Fig.~\ref{fig:fmu}) of radii.}
\end{figure}

The sudden change in behavior is borne out quantitatively by the openness
(Fig.~\ref{fig:openness}\emph{a}), which indicates a sharp increase with
increasing radius around $R=0.55$~nm. Although the range of radii over which
this transition takes place appears to be small ($0.45$~nm to $0.7$~nm) the
cross-sectional area doubles. The maximum number of water molecules actually
found in the cavity in our simulations more than doubles from $21$ to $46$ in
this range of $R$, so that the average environment which each water molecule
experiences changes considerably.

\paragraph{Density}

The radial densities in Fig.~\ref{fig:densities} show  destabilisation of
the liquid phase with decreasing pore radius. Above $R=0.45$~nm distinctive
layering is visible in the pore, and for the larger pores appears as an
extension of the planar layering near the slab. For $R<0.45$~nm no such
features remain and the density is on average close to $0$. The open state can
be identified with liquid water and the closed state with water vapor. In the
continuously open 1~nm-pore, the average density $\langle
n(t)\rangle/n_{\text{bulk}}$ is $0.82$, but $0.032$ in the closed
$0.35$~nm-pore.  \Citet{Brovchenko01} carried out GEMC simulations of the
coexistence of liquid TIP4P water with its vapor in an infinite cylindrical
hydrophobic pore of radius $R=1.075$~nm. At $T=300$~K they obtained a liquid
density of $0.81$ and a vapor density close to $0$, in agreement with the
numbers from our MD simulations.

Analysis of the structure in the radial PMF (Fig.~\ref{fig:radpmf}\emph{b})
lends further support to the above interpretation. Water molecules fill the
narrow pores ($R\lessapprox 0.45$~nm) homogeneously as it is expected for
vapor. For the wider pores, distinct layering is visible as the liquid state
dominates.  The number of layers increases from two to three, with the central
water column being the preferred position initially.  As the radius increases,
the central minimum shifts toward the wall.  For $R=0.7$~nm the center of the
pore is clearly disfavored by $0.2\, k_{B}T$. In the largest pore
($R=1.0$~nm), the influence of curvature on the density already seems to be
negligible as it is almost identical to the situation near a planar
hydrophobic slab.

\paragraph{Kinetics}

Condensation (filling of the pore) and evaporation (emptying) occur in an
avalanche-like fashion as shown in Fig~\ref{fig:wcrossing}\emph{c}.  In our
simulations both events take place within ca.~30~ps, roughly independent of
$R$. However, the actual evaporation and condensation processes seem to follow
different paths, as we can infer from the analysis of the kinetics of the
oscillations.
The time series of Fig.~\ref{fig:states}\emph{a} reveals that the lifetimes of
the open and closed state behave differently with increasing pore radius
(Fig.~\ref{fig:kinetics}\emph{c}): In the range $0.45\ \text{nm} \leq R \leq
0.6\ \text{nm}$, the average time a pore is in the closed state is almost
constant, $\tau_{c} = 1.40\pm0.37$~ns; outside this range no simple functional
relationship is apparent. The average open time can be described as an
exponential $\tau_{o}(R) = a \exp(R/\zeta)$ with $a=1.3\times10^{-5}\ 
\text{ns}$ and $\zeta=4.9\times 10^{-2}\ \text{nm}$ for $0.35\ \text{nm} \leq
R \leq 0.7\ \text{nm}$.

$1/\tau_{o}$ is related to the ``survival probability'' of the liquid state
and $1/\tau_{c}$ to that of the vapor state.  These times characterize the
underlying physical evaporation and condensation processes.  Their very
different dependence on $R$ implies that these processes must be quite
different. The initial condensation process could resemble the evaporation of
water molecules from a liquid surface. Evaporating molecules would not
interact appreciably, so that this process would be rather insensitive to the
area of the liquid-vapor interface $A=\pi R^{2}$ and hence $R$. The disruption
of the liquid pore state, on the other hand, displays very strong dependence
on the radius.  Conceivably, the pore empties once a density fluctuation has
created a vapor bubble that can fill the diameter of the pore and expose the
wall to vapor, its preferred contact phase. The probability for the formation
of a spherical cavity of radius $\lambda$ with exactly $N$ water molecules
inside was determined by \citet{Hummer96}. From their study we find that the
probability $p(\lambda;n)$ for the formation of a bubble of radius $\lambda$
and density below a maximum density $n$ is apparently an exponential.  Once a
bubble with $\lambda \approx R$ develops, the channel rapidly empties but this
occurs with a probability that decreases exponentially with increasing $R$,
which corresponds to the observed exponential increase in $\tau_{o}$. In
particular, for low density bubbles ($n<0.2\, n_{\text{bulk}}$) we estimate
the decay constant in $p(\lambda;n)$ as $2\times 10^{-2}$~nm, which is of the
same order of magnitude as $\zeta$.

From the equilibrium constant $K(R) = T_{c}(R)/T_{o}(R) = \exp[-\beta\, \Delta
F(R)]$ the free energy difference between the two states $\Delta F = F_{c} -
F_{o}$ can be calculated. $\Delta F$ increases linearly with the pore radius
(inset of Fig.~\ref{fig:openness}\emph{a}), $ \beta\, \Delta F(R) = a_{0} +
a_{1} R$ with $a^{\text{kin}}_{0}=-13.2\pm1.4$ and
$a^{\text{kin}}_{1}=23.7\pm3.0\ \text{nm}^{-1}$.  Together with $K(R)$, the
gating behavior of the pore is characterized \citep{Sakmann83}. In this
sense, the MD calculations have related the input structure to a
``physiological'' property of the system. (Note, however, that the time scales
of ion channel gating and of the oscillations observed here differ by five
orders of magnitude.)

\paragraph{Free energy density}

\begin{figure}
  \centering
  \includegraphics[clip]{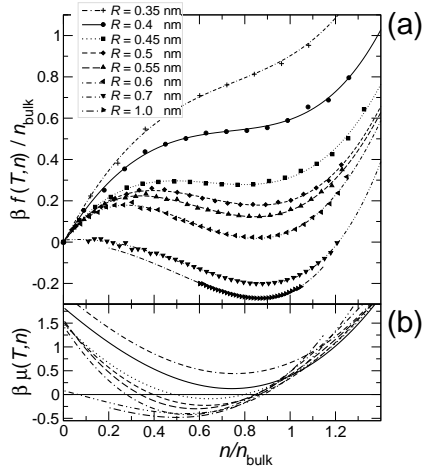}
  \caption{\label{fig:fmu}%
    (\emph{a}) Free energy density $f(T,n)$ at constant $T=300$~K. (\emph{b})
    Chemical potential $\mu(T,n)$. $n$ is the water density in the pore,
    normalized to $n_{\text{bulk}}=53.7$~mol\,l$^{-1}$. $f$ is given in units
    of $k_{B}T$ and the inverse of the liquid molecular volume of bulk water
    ($v_{l}^{-1}=n_{\text{bulk}}$).
    Two minima correspond to the observed two-state behavior. The vapor state
    becomes metastable with increasing radius and for $R>0.55$~nm the liquid
    state is globally stable.
    $f(T,n;R=1.0\ \text{nm})$ is drawn with an arbitrary offset.
  }
\end{figure}

The Helmholtz free energy density $f(T,n;R)$ displays one or two minima
(Fig.~\ref{fig:fmu}\emph{a}): one for the empty pore ($n=0$) and one in the
vicinity of the bulk density. The 0.45~nm pore is close to a transition point
in the free energy landscape: the minimum for the filled pore is very shallow
and disappears at smaller radii ($R=0.4$~nm and 0.35~nm).  For very large and
very small radii, only one thermodynamic stable state exists: liquid or vapor.
For intermediate radii, a metastable state appears.  Near $R=0.55$~nm both
states are almost equally probable although they do not coexist spatially
because the pore is finite and small. In infinite pores \emph{spatially}
alternating domains of equal length would be expected \citep{Privman83} and
were actually observed in MD simulations \citep{Peterson88}. The oscillating
states in short pores, on the other hand, alternate \emph{temporally}, thus
displaying a kind of ``time-averaged'' coexistence.
%
For higher densities $n/n_{\text{bulk}}>1$ the curves start to resemble
parabolas, similar to a parabolic $f(T,n)$ seen for cylindrical volumes (data
not shown) and spherical cavities \citep{Hummer96} in bulk water.

The chemical potential (Fig.~\ref{fig:fmu}\emph{b}) shows the transition from
the stable vapor state, $\mu(T,n)>0$, through the two-state regime to the
stable liquid state, $\mu(T,n)<0$. The features of $\mu(T,n)$ indicate that
the condensation (and evaporation) processes occur in an avalanche-like
fashion: Let the density in the pore be at the transition state, the left zero
of $\mu$. If the density is perturbed to increase slightly then $\mu$ becomes
negative. Every additional molecule added to the pore decreases the free
energy further by an amount $\mu$ while the increase in density lowers the
chemical potential even more.  This leads to the avalanche of condensation. It
only stops when the stable state, the right zero of $\mu$, is reached. Now a
further addition of molecules to the pore would actually increase the free
energy and drive the system back into the stable state.  Similarly, a
perturbation that decreases the density in the transition state leads to
accelerated evaporation.

From the probability distribution $p(N)$ the free energy difference between
closed and open state $\Delta F(R)$ is calculated,
$a^{\text{eq}}_{0}=-14.9\pm2.2$ and $a^{\text{eq}}_{1}=26.3\pm4.1\ 
\text{nm}^{-1}$, consistent with the estimate from the kinetics. $\Delta F(R)$
(inset of Fig.~\ref{fig:openness}\emph{a}) shows the transfer of stability
from the vapor state for small $R$ to the liquid state for large $R$. The
coexistence regime is at $\Delta F(R_{c}=0.57\ \text{nm})=0$.

\paragraph{Dynamics}

MD simulations not only allow us to investigate the thermodynamic properties
of the system but also the dynamical behavior of individual molecules. A few
selected water molecules are depicted in Fig.~\ref{fig:wcrossing}\emph{c}
shortly before the pore empties. They show a diverse range of behaviors and no
single-file like motion of molecules is visible in the liquid state. On
evaporation (and condensation) the state changes within ca.\ 30~ps.

\begin{table}[tb]
  \caption{
      Dynamical properties of water in hydrophobic pores. %
      $R$ is the van der Waals pore radius,
      with $R=\infty$ denoting the bulk. $\langle\omega\rangle$ is the
      openness. The mean permeation
      time $\langle\tau_{p}\rangle$ is measured relative to the bulk value,
      $\langle \tau_{p,\text{bulk}} \rangle = 29.9\pm0.1$~ps. The equilibrium
      current density  $j_{0}$ is the total number of permeant water molecules
      per unit time and unit 
      area ($j_{0,\text{bulk}}=320\pm3~$ns$^{-1}\,\text{nm}^{-2}$).  The
      diffusion coefficient along the pore axis $D_z$ is normalized to the
      bulk value of SPC water at 300~K and 1~bar
      ($D_{\text{bulk}}=4.34\pm0.01\ \text{nm}^{2}\, \text{ns}^{-1}$). One
      standard deviation errors in the last decimals are given in parentheses.}
  \label{tab:dynamics}
  \begin{center}
    \begin{tabular}{r@{.}lr@{.}lr@{.}lr@{.}lr@{.}l}
      \toprule
      \multicolumn{2}{c}{$R/\text{nm}$}  &
      \multicolumn{2}{c}{$\langle\omega\rangle$} &
      \multicolumn{2}{c}{$\langle\tau_p\rangle/\langle\tau_{p,\text{bulk}}\rangle$}&   
      \multicolumn{2}{c}{$j_{0}/j_{0,\text{bulk}}$} &
      \multicolumn{2}{c}{$D_z/D_\text{bulk}$}  \\
      \midrule
      0&35 & 0&008(2) & 0&482(61) & 0&025(2) & \multicolumn{2}{c}{} \\
      0&4  & 0&015(5) & 0&421(25) & 0&027(2) &  2&87(9)\\
      0&45 & 0&101(30) & 0&629(12) & 0&109(3) &  2&27(4)\\
      0&5  & 0&181(41) & 0&729(10) & 0&194(4) &  1&91(3)\\
      0&55 & 0&291(89) & 0&786(8) & 0&279(4) &  1&87(3)\\
      0&6  & 0&775(79) & 0&833(5) & 0&721(6) &  1&32(1)\\
      0&7  & 0&999(1) & 0&799(3) & 1&004(7) &  1&25(0)\\
      1&0  & 1&000(0) & 0&819(2) & 1&011(5) &  1&18(0)\\         
      \multicolumn{2}{l}{$\infty$} & \multicolumn{2}{c}{} 
                  &     1&000(3) &     1&000(8) &  1&00(0)\\
      \bottomrule                         
    \end{tabular}
  \end{center}
\end{table}

The mean permeation time $\langle\tau_{p}\rangle$ in Table~\ref{tab:dynamics}
increases with the pore radius, i.e.\ water molecules permeate narrow
hydrophobic pores faster than they diffuse the corresponding distance in bulk
water (the reference value).  This is consistent with higher diffusion
coefficients $D_{z}$ in the narrow pores (up to almost three times the bulk
value). The diffusion coefficient perpendicular to the pore axis, $D_{xy}$,
drops to approximately half the bulk value.  \Citet{Marti01} also observe
increased diffusion in simulations on water in carbon nanotubes ($D_{z}\leq
1.65\, D_{\text{bulk}}$) and a corresponding decrease in $D_{xy}$.
Experimental studies on water transport through desformyl gA \citep{Saparov00}
can be interpreted in terms of a $D_{z}$ of five times the bulk value.
Histograms (data not shown) for $\tau_{p}$ show that there is a considerable
population of `fast' water molecules (e.g.\ the black and the dark gray one in
Fig.~\ref{fig:wcrossing}\emph{c}) with $\tau_{p}$ between 2 and 10~ps, which
is not seen in bulk water.  The acceleration of water molecules in the pore
can be understood as an effect of 1D confinement. The random 3D motion is
directed along the pore axis and the particle advances in this direction
preferentially. The effect increases with decreasing radius, i.e.\ increasing
confinement. The average equilibrium current density $j_{0}$ follows the trend
of the openness closely but more detailed time-resolved analysis shows water
translocation to occur in bursts in all pores. In narrow pores, bursts
occurring during the ``closed'' state contribute up to 77\% of the total flux
(data not shown).  For single-file pores, simulations \citep{Hum01,DeGroot02}
and theory \citep{Berezhkovskii02} also point towards concerted motions as the
predominant form of transport.

\paragraph{Capillary condensation}

The behavior as described so far bears the hallmarks of capillary
condensation and evaporation \citep{Rowlinson82,Evans90,Gelb99} although it is
most often associated with physical systems which are macroscopically extended
in at least one dimension such as slits or long pores. Capillary condensation
can be discussed in terms of the Kelvin equation \citep{Christenson01},
\begin{equation}
  \label{eq:Kelvin}
  \ln \frac{p}{p_{0}} = -\frac{\beta \gamma_{lv} v_{l}}{r},
\end{equation}
which describes how vapor at pressure $p$ relative to its bulk-saturated
pressure $p_{0}$ coexists in equilibrium with its liquid.  Liquid and vapor
are divided by an interfacial meniscus of curvature $1/r$ ($r>0$ if the
surface is convex); $\gamma_{lv}$ is the surface tension between liquid and
vapor and $v_{l}$ the molecular volume of the liquid. Although the Kelvin
equation is not expected to be quantitative in systems of dimensions of only a
few molecular diameters it is still useful for obtaining a qualitative
picture.  Curvature $1/r$ and contact angle $\theta$ in a cylindrical pore of
radius $R$ are related by $R=r\cos\theta$. With Young's equation,
$\gamma_{wv}=\gamma_{wl}+\gamma_{lv}\cos\theta$, Eq.~\ref{eq:Kelvin} becomes
\begin{equation}
  \label{eq:kelvinR}
  \ln \frac{p(R)}{p_{0}} = -\frac{\beta (\gamma_{wv}-\gamma_{wl}) v_{l}}{R},
\end{equation}
independent of the interface. For our system, the surface tension between
liquid water and the wall, $\gamma_{wl}>0$, and between vapor and the wall,
$\gamma_{wv}>0$, are fixed quantities.  The hydrophobicity of the wall implies
$\gamma_{wv}<\gamma_{wl}$, i.e.\ the wall is preferentially in contact with
vapor; $v_{l}$ can be considered constant.  Hence, for a given pore of radius
$R$ there exists one vapor pressure $p(R)>p_{0}$ at which vapor and liquid can
exist in equilibrium. Water only condenses in the pore if the actual vapor
pressure exceeds $p(R)$. Otherwise, only vapor will exist in the pore. The
effect is strongest for very narrow pores. Hence a higher pressure is required
to overcome the surface contributions, which stabilize the vapor phase in
narrow pores.  The pressure fluctuates locally in the liquid bulk
``reservoir\@.'' These fluctuations can provide an increase in pressure above
the saturation pressure in the pore and thus drive oscillations between vapor
and liquid.

\paragraph{Comparison with experiments, simulations, and a theoretical model}

Experiments on aquaporins \citep{Yang97,Pohl01} and gA
\citep{Pohl00,Saparov00} yield osmotic permeability coefficients of water,
$p_{f}$, of the order of $10^{-12}$ to $10^{-14}\ \text{cm}^{3}
\text{s}^{-1}$. We calculate $p_{f} = \frac{1}{2} \Phi_{0} v_{l}$ from the
equilibrium flux of our MD simulations \citep{DeGroot02} and find that narrow
($R=0.35$~nm and $0.4$~nm), predominantly ``closed'' pores have $p_{f}\approx
5\times 10^{-14}\ \text{cm}^{3} \text{s}^{-1}$, that is, the same magnitude as
Aqp1, AqpZ, and gA (see Table~\ref{tab:permeabilities}).
\begin{table}
  \centering
  \caption{Osmotic permeability coefficient $p_{f}$ and equilibrium flux
    $\Phi_{0}$ of water in selected simulations and experiments. We used 
    the relationship $p_{f} = \frac{1}{2} \Phi_{0} v_{l}$
    from Ref.~\protect\citealp{DeGroot02} 
    in order to  compare non-equilibrium experiments (upper half of the table)
    with   equilibrium   molecular dynamics simulations (lower half). $v_{l} =
    3.09\times 10^{-23}$~cm$^{3}$ is 
    the volume of a water molecule in the liquid state.}
  \label{tab:permeabilities}
  \begin{minipage}{\linewidth}
    \centering
    \renewcommand{\footnoterule}{}
    \renewcommand{\thefootnote}{\thempfootnote}
  \begin{tabular}{lcr@{}lr@{}l}
    \toprule
        & Ref.
        & \multicolumn{2}{c}{$p_{f}\times10^{14}$}
        & \multicolumn{2}{c}{$\Phi_{0}$} \\
        &
        & \multicolumn{2}{c}{[$\text{cm}^{3} \text{s}^{-1}$]}
        & \multicolumn{2}{c}{[$\text{ns}^{-1}$]} \\
    \midrule
    Aqp1 & \citep{Yang97}    &  4&.9   &   3&.2\\
    Aqp4 & \citep{Yang97}    & 15&     &   9&.7\\
    AqpZ & \citep{Pohl01}    &  2&.0   &   1&.3\\
    gA\footnote{bacterial peptide channel gramicidin A} %
         & \citep{Pohl00}    &  1&.6   &   1&.0\\
    desformyl gA\footnote{desformylated gramicidin A} %
         & \citep{Saparov00} & 110&    &  71&  \\
    \midrule
    $R=0.35$~nm & & 4&.0 & 2&.6\\
    $R=0.40$~nm & & 5&.7 & 3&.7\\
    $R=0.45$~nm & & 30&.0  & 19&.4 \\
    $R=0.50$~nm & & 66&.5  & 43&.0 \\
    $R=0.55$~nm & & 117& & 75&.8 \\
    $R=0.60$~nm & & 363& & 235& \\
    $R=0.70$~nm & & 700& & 453& \\
    $R=1.0$~nm  & & 1480& & 956\\
    carbon nanotube\footnote{$(6,6)$ carbon nanotube, $R\approx0.24$~nm} %
                  & \citep{Hum01}     & 26&.2 & 16&.9\\
    desformyl gA (DH)\footnote{desformyl gA in the double-helical
    conformation} %
                  & \citep{DeGroot02} & 10&   &  5&.8\\ 
    \bottomrule
  \end{tabular}
  \end{minipage}
\end{table}
As these pores are longer (ca.\ $2$~nm) and narrower ($R<0.2$~nm) than our
model pores, strategically placed hydrophilic groups \citep{Tajkhorshid02}
seem to be needed to stabilize the liquid state and facilitate water transport
in these cases.

Recently \citet{Giaya02} presented an analytical mean-field model for water in
infinite cylindrical hydrophobic micropores. They predict the existence of a
critical radius $R_{c}$ for the transition from a thermodynamically stable
water vapor phase to a liquid phase. The crucial parameter that $R_{c}$
depends on is the water-wall interaction.  We choose the effective fluid-wall
interaction $\epsilon_{\text{eff}}=\rho_{w}\, \epsilon_{fw}$, the product of
the density of wall atoms with the well-depth of the fluid-wall interaction
potential, as a parameter to compare different simulations because this seems
to be the major component in the analytical fluid-wall interaction.
\begin{table}[tb]
    \caption{Comparison of different studies of water in hydrophobic
      pores. The wall-atom density $\rho_{w}$ is in units of nm$^{-3}$,
      the fluid-wall interaction $\epsilon_{fw}$ in kJ\,mol$^{-1}$ and
      the effective  interaction strength $\epsilon_{\text{eff}}$ in
      kJ\,mol$^{-1}$\,nm$^{-3}$.  The critical pore radius $R_{c}$ is given in
      nm. The pore length was 0.8~nm in this work, 1.7~nm in the
      carbon nanotube simulations \citep{Hum01} and infinite in the mean field
      model  \citep{Giaya02}.} 
    \label{tab:comparison}
    \begin{center}
    \begin{tabular}{crr@{.}lrc}
      \toprule
      Ref.   &  $\rho_{w}$ & \multicolumn{2}{c}{$\epsilon_{fw}$} 
                           & $\epsilon_{\text{eff}}$ 
                           & \multicolumn{1}{c}{$R_{c}$}\\
      \midrule
      this work       &   8 & 0&906493 &  7 & $\approx 0.57$\\
      \citep{Hum01}   &  50 & 0&478689 & 24 & $<0.24$\\
                      &     & 0&272937 & 14 & $>0.24$\\
      \citep{Giaya02} & 110 & \multicolumn{2}{l}{0} 
                                       & 0  &  1500\\
                      &     & 1&4      & 154 & 190\\
                      &     & 1&45     & 160 & 0.35\\
                      &     & 2&0      & 220 & 0\\
      \bottomrule                      
    \end{tabular}
  \end{center}
\end{table}
As shown in Table~\ref{tab:comparison}, compared to carbon nanotube
simulations our pore has a very small $\epsilon_{\text{eff}}$ and thus can be
considered extremely hydrophobic. This explains why \citet{Hum01} observe
permanently water filled nanotubes with a radius of only 0.24~nm although
their bare fluid-wall interaction potential is weaker than in our model. The
much higher density of wall atoms in the nanotube, however, more than
mitigates this. Once they lower their $\epsilon_{\text{eff}}$ to double of our
value, they also observe strong evaporation. This suggests that the close
packing of wall atoms within a nanotube may result in behavior not seen in
biological pores. The mean field model agrees qualitatively with the
simulations as it also shows a sharp transition and high sensitivity to
$\epsilon_{\text{eff}}$.

\section{Conclusions}
\label{sec:conclusions}

We have described oscillations between vapor and liquid states in short
($L_{P}=0.8$~nm), hydrophobic pores of varying radius ($0.35$~nm$\leq R \leq
1.0$~nm).  Qualitatively, this behavior is explained as capillary evaporation,
driven by pressure/density fluctuations in the water ``reservoir'' outside the
pore.
Similar behavior is found in simulations by different authors with different
water models [SPC (this work), SPC/E, TIP3P (data not shown), TIP3P
\citep{Hum01}, TIP4P \citep{Brovchenko02}] in different nanopores [atomistic
flexible models (this work), carbon nanotubes \citep{Hum01}, spherical
cavities \citep{Bro00} and smooth cylinders \citep{Brovchenko02,Allen02}].

We presented a radically simplified model for a nanopore that is perhaps more
hydrophobic than in real proteins [although we note the existence of a
hydrophobic pore in the MscS channel \citep{Bass02}]. From comparison with
experimental data on permeability we conclude that strategically placed
hydrophilic groups are essential for the functioning of protein pores. The
comparatively high permeability of our ``closed'' pores suggests pulsed water
transport as one possible mechanism in biological water pores.  Local
hydrophobic environments in pores may promote pulsatory collective transport
and hence rapid water and solute translocation.

Our results indicate new, intrinsically collective dynamic behavior not seen
on simulation time scales currently considered sufficient in biophysical
simulations. These phase oscillations in simple pores---a manifestation of the
hydrophobic effect---require more than 50~ns of trajectory data to yield a
coherent picture over a free energy range of $6\,k_{B}T$. We thus cannot
safely assume that the behavior of water within complex biological pores may
be determined by extrapolation from our knowledge of the bulk state or short
simulations alone.

\paragraph{Acknowledgments}
This work was funded by The Wellcome Trust. Our thanks to all of our
colleagues for their interest in this work, especially Joanne Bright, Jos\'e
Faraldo-G\'omez, Andrew Horsfield and Richard Law.

%
%

%


\begin{thebibliography}{42}

\bibitem[Doyle et~al.(1998)Doyle, Morais-Cabral, Pf{\"u}tzner, Kuo, Gulbis,
  Cohen, Chait, and MacKinnon]{Doy98}
Doyle, D.~A., Morais-Cabral, J., Pf{\"u}tzner, R.~A., Kuo, A., Gulbis, J.~M.,
  Cohen, S.~L., Chait, B.~T., \& MacKinnon, R. (1998) {\em Science}
  \textbf{280}, 69--77.

\bibitem[Chang et~al.(1998)Chang, Spencer, Lee, Barclay, and Rees]{Cha98}
Chang, G., Spencer, R.~H., Lee, A.~T., Barclay, M.~T., \& Rees, D.~C. (1998)
  {\em Science} \textbf{282}, 2220--2226.

\bibitem[Fu et~al.(2000)Fu, Libson, Miercke, Weitzman, Nollert, Krucinski, and
  Stroud]{Fu00}
Fu, D., Libson, A., Miercke, L.~J., Weitzman, C., Nollert, P., Krucinski, J.,
  \& Stroud, R.~M. (2000) {\em Science} \textbf{290}, 481--486.

\bibitem[Sui et~al.(2001)Sui, Han, Lee, Walian, and Jap]{Sui01}
Sui, H., Han, B.~G., Lee, J.~K., Walian, P., \& Jap, B.~K. (2001) {\em Nature}
  \textbf{414}, 872--878.

\bibitem[Dutzler et~al.(2002)Dutzler, Campbell, Cadene, Chait, and
  MacKinnon]{Dutzler02}
Dutzler, R., Campbell, E.~B., Cadene, M., Chait, B.~T., \& MacKinnon, R. (2002)
  {\em Nature} \textbf{415}, 287--294.

\bibitem[Bass et~al.(2002)Bass, Strop, Barclay, and Rees]{Bass02}
Bass, R.~B., Strop, P., Barclay, M., \& Rees, D.~C. (2002) {\em Science}
  \textbf{298}, 1582--1587.

\bibitem[Hille(2001)]{Hille01}
Hille, B. (2001) {\em Ion Channels of Excitable Membranes} (Sinauer Associates,
  Sunderland MA, U.S.A.), 3rd ed.

\bibitem[Tieleman et~al.(2001)Tieleman, Biggin, Smith, and Sansom]{Tieleman01}
Tieleman, D.~P., Biggin, P.~C., Smith, G.~R., \& Sansom, M. S.~P. (2001) {\em
  Quart. Rev. Biophys.} \textbf{34}, 473--561.

\bibitem[Yang et~al.(1997)Yang, van Hoek, and Verkman]{Yang97}
Yang, B., van Hoek, A.~N., \& Verkman, A.~S. (1997) {\em Biochemistry}
  \textbf{36}, 7625--7632.

\bibitem[Pohl et~al.(2001)Pohl, Saparov, Borgnia, and Agre]{Pohl01}
Pohl, P., Saparov, S.~M., Borgnia, M.~J., \& Agre, P. (2001) {\em
  Proc.~Natl.~Acad.~Sci.~USA} \textbf{98}, 9624--9629.

\bibitem[Fujiyoshi et~al.(2002)Fujiyoshi, Mitsuoka, de~Groot, Philippsen,
  Grubm{\"u}ller, Agre, and Engel]{Fujiyoshi02R}
Fujiyoshi, Y., Mitsuoka, K., de~Groot, B.~L., Philippsen, A., Grubm{\"u}ller,
  H., Agre, P., \& Engel, A. (2002) {\em Curr.~Opin.~Struct.~Biol.}
  \textbf{12}, 509--515.

\bibitem[Pohl and Saparov(2000)]{Pohl00}
Pohl, P. \& Saparov, S.~M. (2000) {\em Biophys.~J.} \textbf{78}, 2426--2434.

\bibitem[Saparov et~al.(2000)Saparov, Antonenko, , and Pohl]{Saparov00}
Saparov, S.~M., Antonenko, Y.~N., , \& Pohl, P. (2000) {\em Biophys.~J.}
  \textbf{79}, 2526--2534.

\bibitem[de~Groot et~al.(2002)de~Groot, Tieleman, Pohl, and
  Grubm{\"u}ller]{DeGroot02}
de~Groot, B.~L., Tieleman, D.~P., Pohl, P., \& Grubm{\"u}ller, H. (2002) {\em
  Biophys.~J.} \textbf{82}, 2934--42.

\bibitem[Tajkhorshid et~al.(2002)Tajkhorshid, Nollert, Jensen, Miercke,
  O'Connell, Stroud, and Schulten]{Tajkhorshid02}
Tajkhorshid, E., Nollert, P., Jensen, M.~{\O}., Miercke, L.~J., O'Connell, J.,
  Stroud, R.~M., \& Schulten, K. (2002) {\em Science} \textbf{296}, 525--530.

\bibitem[de~Groot and Grubm{\"u}ller(2001)]{DeGroot01}
de~Groot, B.~L. \& Grubm{\"u}ller, H. (2001) {\em Science} \textbf{294},
  2353--2357.

\bibitem[Finkelstein(1987)]{Finkelstein87}
Finkelstein, A. (1987) {\em Water Movement Through Lipid Bilayers, Pores, and
  Plasma Membranes. Theory and Reality} (John Wiley \& Sons, New York).

\bibitem[Christenson(2001)]{Christenson01}
Christenson, H.~K. (2001) {\em J.~Phys.: Condens.~Matter} \textbf{13},
  R95--R133.

\bibitem[Gelb et~al.(1999)Gelb, Gubbins, Radhakrishnan, and
  Sliwinska-Bartkowiak]{Gelb99}
Gelb, L.~D., Gubbins, K.~E., Radhakrishnan, R., \& Sliwinska-Bartkowiak, M.
  (1999) {\em Rep.~Prog.~Phys.} \textbf{62}, 1573--1659.

\bibitem[Lynden-Bell and Rasaiah(1996)]{Lyn96}
Lynden-Bell, R.~M. \& Rasaiah, J.~C. (1996) {\em J.~Chem.~Phys.} \textbf{105},
  9266--9280.

\bibitem[Allen et~al.(1999)Allen, Kuyucak, and Chung]{All99}
Allen, T.~W., Kuyucak, S., \& Chung, S.-H. (1999) {\em J.~Chem.~Phys.}
  \textbf{111}, 7985--7999.

\bibitem[Hummer et~al.(2001)Hummer, Rasaiah, and Noworyta]{Hum01}
Hummer, G., Rasaiah, J.~C., \& Noworyta, J.~P. (2001) {\em Nature}
  \textbf{414}, 188--190.

\bibitem[Beckstein et~al.(2001)Beckstein, Biggin, and Sansom]{Beckstein01}
Beckstein, O., Biggin, P.~C., \& Sansom, M. S.~P. (2001) {\em J.~Phys.~Chem.~B}
  \textbf{105}, 12902--12905.

\bibitem[Brovchenko et~al.(2000)Brovchenko, Paschek, and Geiger]{Bro00}
Brovchenko, I., Paschek, D., \& Geiger, A. (2000) {\em J.~Chem.~Phys.}
  \textbf{113}, 5026--5036.

\bibitem[Brovchenko and Geiger(2002)]{Brovchenko02}
Brovchenko, I. \& Geiger, A. (2002) {\em J.~Mol.~Liquids} \textbf{96--97},
  195--206.

\bibitem[Unwin(2000)]{Unw00}
Unwin, N. (2000) {\em Phil.~Trans.~Roy.~Soc.~London~B} \textbf{355},
  1813--1829.

\bibitem[Lindahl et~al.(2001)Lindahl, Hess, and van~der Spoel]{Lindahl01}
Lindahl, E., Hess, B., \& van~der Spoel, D. (2001) {\em J. Mol. Mod.}
  \textbf{7}, 306--317.
\newblock {http://www.gromacs.org}.

\bibitem[Hermans et~al.(1984)Hermans, Berendsen, van Gunsteren, and
  Postma]{Hermans84}
Hermans, J., Berendsen, H. J.~C., van Gunsteren, W.~F., \& Postma, J. P.~M.
  (1984) {\em Biopolymers} \textbf{23}, 1513--1518.

\bibitem[Darden et~al.(1993)Darden, York, and Pedersen]{Dar93}
Darden, T., York, D., \& Pedersen, L. (1993) {\em J.~Chem.~Phys.} \textbf{98},
  10089--10092.

\bibitem[Berendsen et~al.(1984)Berendsen, Postma, DiNola, and Haak]{Ber84}
Berendsen, H. J.~C., Postma, J. P.~M., DiNola, A., \& Haak, J.~R. (1984) {\em
  J.~Chem.~Phys.} \textbf{81}, 3684--3690.

\bibitem[Sakmann and Neher(1983)]{Sakmann83}
Sakmann, B. \& Neher, E., eds. (1983) {\em Single-Channel Recordings} (Plenum
  Press, New York).

\bibitem[Preusser(1989)]{Preusser89}
Preusser, A. (1989) {\em ACM Trans.~Math.~Softw.} \textbf{15}, 79--89.
\newblock {http://www.fhi-berlin.mpg.de/grz/pub/xfarbe/}.

\bibitem[Brovchenko et~al.(2001)Brovchenko, Geiger, and
  Oleinikova]{Brovchenko01}
Brovchenko, I., Geiger, A., \& Oleinikova, A. (2001) {\em
  Phys.~Chem.~Chem.~Phys.} \textbf{3}, 1567--1569.

\bibitem[Hummer et~al.(1996)Hummer, Garde, Garc{\'\i}a, Pohorille, and
  Pratt]{Hummer96}
Hummer, G., Garde, S., Garc{\'\i}a, A.~E., Pohorille, A., \& Pratt, L.~R.
  (1996) {\em Proc.~Natl.~Acad.~Sci.~USA} \textbf{93}, 8951--8955.

\bibitem[Privman and Fisher(1983)]{Privman83}
Privman, V. \& Fisher, M.~E. (1983) {\em J. Stat. Phys.} \textbf{33}, 385--417.

\bibitem[Peterson et~al.(1988)Peterson, Gubbins, Heffelfinger, Marconi, and van
  Smol]{Peterson88}
Peterson, B.~K., Gubbins, K.~E., Heffelfinger, G.~S., Marconi, U. M.~B., \& van
  Smol, F. (1988) {\em J.~Chem.~Phys.} \textbf{88}, 6487--6500.

\bibitem[Mart{\'\i} and Gordillo(2001)]{Marti01}
Mart{\'\i}, J. \& Gordillo, M.~C. (2001) {\em Phys.~Rev.~E} \textbf{64},
  021504--1.

\bibitem[Berezhkovskii and Hummer(2002)]{Berezhkovskii02}
Berezhkovskii, A. \& Hummer, G. (2002) {\em Phys.~Rev.~Lett.} \textbf{89},
  065403{--}1--4.

\bibitem[Rowlinson and Widom(1982)]{Rowlinson82}
Rowlinson, J.~S. \& Widom, B. (1982) {\em Molecular Theory of Capillarity}
  (Clarendon Press, Oxford).

\bibitem[Evans(1990)]{Evans90}
Evans, R. (1990) {\em J.~Phys.: Condens.~Matter} \textbf{2}, 8989--9007.

\bibitem[Giaya and Thompson(2002)]{Giaya02}
Giaya, A. \& Thompson, R.~W. (2002) {\em J.~Chem.~Phys.} \textbf{117},
  3464--3475.

\bibitem[Allen et~al.(2002)Allen, Melchionna, and Hansen]{Allen02}
Allen, R., Melchionna, S., \& Hansen, J.-P. (2002) {\em Phys.~Rev.~Lett.}
  \textbf{89}, 175502--1--175502--4.
\end{thebibliography}
\end{document}